\documentclass{aa}  
\usepackage{keyval}
\usepackage[switch]{lineno} 
\usepackage{graphicx}

\usepackage{txfonts}
\usepackage{hyperref}

\begin{document} 
\nolinenumbers       

   \title{Formation of millisecond pulsar-helium star binaries}

   \author{Zhu-Ling Deng\inst{1,2,3}\and Xiang-Dong Li\inst{1,2\thanks{Corresponding author. Email: lixd@nju.edu.cn}}\and Yong Shao\inst{1,2}\and Ying-Han Mao\inst{1,2}\and Long Jiang\inst{4}
          }

   \institute{School of Astronomy and Space Science, Nanjing University, Nanjing 210023, China
         \and
             Key Laboratory of Modern Astronomy and Astrophysics (Nanjing University), Ministry of Education, Nanjing 210023, China
        \and
        Yunnan Observatories, Chinese Academy of Sciences, Kunming 650216, China
        \and 
            School of Science, Qingdao University of Technology, Qingdao 266525, China
             }



  \abstract
   {PSR J1928+1815, the first recycled pulsar-helium (He) star binary discovered by the Five-hundred-meter Aperture Spherical radio Telescope, consists of a 10.55 ms pulsar and a companion star with mass $1-1.6\,M_{\sun}$ in a 0.15-day orbit. Theoretical studies suggest that this system originated from a neutron star (NS) intermediate-mass or high-mass X-ray binary that underwent common envelope (CE) evolution,  leading to the successful ejection of the giant envelope. The traditional view is that hypercritical accretion during the CE phase may have recycled the NS. However, the specific mechanism responsible for  accelerating its spin period remains uncertain due to the complex processes involved in CE evolution.}
   {In this study, we investigate the influence of Roche lobe overflow (RLO) accretion that takes place prior to the CE phase on the spin evolution of NSs. Our primary objective is to clarify how this process affects the spin characteristics of pulsars.}
   {We utilized the stellar evolution code \texttt{MESA} and the binary population synthesis code \texttt{BSE} to model the formation and evolution of NS-He star binaries. We calculated the distributions of the orbital period, He star mass, NS spin period, and magnetic field for NS + He star systems in the Galaxy.}
   {Our results indicate that RLO accretion preceding the CE phase could spin up NSs to millisecond periods through super-Eddington accretion. Considering a range of CE efficiencies $\alpha_{\rm CE}$ from 0.3 to 3, we estimate the birthrate (total number) of NS + He star systems in our Galaxy to be 5.2$\times 10^{-5}$ yr$^{-1}$ (626 systems) to 1.9$\times 10^{-4}$ yr$^{-1}$ (2684 systems).}
   {}


   \maketitle
%

\section{Introduction}

The recent discovery of the pulsar J1928+1815 by the Five-hundred-meter Aperture Spherical radio Telescope (FAST) offers a novel perspective on the formation and evolution of millisecond pulsars \citep[MSPs,][]{Yang2025}. This pulsar, with a spin period of 10.55 ms, resides in a binary system with an orbital period of approximately 0.15 day. Its companion is thought to be a helium (He) star with  mass $1-1.6\,M_\odot$.  

Observationally, binary systems containing stripped He stars have been identified in the Magellanic Clouds \citep{Drout2023}. In the Milky Way, X-ray binaries with He star donors have also been discovered, with accretors including white dwarfs \citep[WDs,][]{Mereghetti2009} and neutron stars \citep[NSs,][]{Mereghetti2016}. However, prior to this discovery, no binary systems containing recycled pulsars and He star companions had been identified. The detection of PSR J1928+1815 fills this observational gap, representing the first confirmed binary system of this kind. It provides crucial evidence for testing and extending existing binary evolution theories and opens a new window for in-depth studies of the accretion-induced spin-up process in MSPs. Through detailed observations and theoretical analyses of this system, we can gain a better understanding of the formation mechanisms and physical properties of MSPs.

MSPs are widely believed to form through the recycling of old NSs via accretion of matter and angular momentum  from a companion star in low-mass X-ray binaries \citep[LMXBs,][]{Alpar1982,Radhakrishnan1982}. This hypothesis is supported by the detection of millisecond X-ray pulsations in LMXBs and observations of transitions from accreting millisecond X-ray pulsars to MSPs \citep{Wijnands1998,Archibald2009,Papitto2013,Roy2015}. 

In contrast, systems such as PSR J1928+1815 are thought to originate from intermediate-mass or high-mass X-ray binaries (IMXBs or HMXBs) consisting of a NS and a more massive companion star \citep[e.g.,][]{Vigna2018,Shao2018a,Deng2024a}. When the companion star evolves to be a giant star and overflows its Roche-lobe, mass transfer (MT) will typically be dynamically unstable because of the large mass ratio. The binary enters the common envelope (CE) phase \citep{Ivanova2013}, during which the giant star's envelope is shed. If the binary survives CE evolution, a recycled NS and a He star, which is the core of the giant star, are left. 

Theoretically, accelerating a NS's spin period to 10 ms requires accretion of at least $0.01\,M_\odot$  of matter \citep[e.g.,][]{Tauris2012}, and a NS in intermediate-mass or high-mass X-ray binaries may undergo multiple accretion phases. Initially, there is accretion from the stellar wind of the giant companion before Roche lobe overflow (RLO). While wind accretion can provide some accreted matter \citep[$\sim0.005\,M_{\odot}$,][]{Yang2025}, its impact on the NS's spin is uncertain, as it may both accelerate and decelerate the spin \citep[see][and references therein]{Tauris2023,Mao2024}. Current observations of wind-fed X-ray pulsars indicate that the spin periods of NSs range from $\sim 1-10,000$ s \citep[see][Figure 1]{Wengss2024}, with a typical value of $\sim 100$ s \citep{Lyne2012}. 

Subsequently, there is the RLO phase. During this phase, if the MT rate is constrained by the Eddington limit, accreting $0.01\,M_\odot$
would necessitate a sustained MT time of over $5\times 10^5$ years. Finally, there is the CE evolution phase. Theoretical studies suggest that the NS could experience hypercritical accretion (>$2 \times 10^{-4}\,M_{\odot}$\,yr$^{-1}$) in the CE phase due to the loss of accretion energy through neutrino emission \citep{Houck1991,Ivanova2011}. More recent three-dimensional hydrodynamical simulations indicate that the NS can accrete at most a few percent of $1\,M_\odot$ due to density gradients in the CE structures \citep{MacLeod2015a,MacLeod2015b}. Furthermore, accretion in CE environments typically follows the spherically symmetric Hoyle-Lyttleton model, with no persistent accretion disks forming around the embedded objects \citep{MacLeod2017,Hutchinson2024,Rosselli2024}. This aligns with the significant misalignment between the spin and orbital axes observed in PSR J1928+1815 \citep{Yang2025}, suggesting inefficient angular momentum transfer during CE evolution.

In this work we focus on the RLO process proceeding CE evolution, motivated by the observational evidence for super-Eddington accretion onto magnetized NSs in intermediate-mass or high-mass X-ray binaries. Several pulsating ultraluminous X-ray sources (ULXs) have been discovered \citep[][for a review]{Kuranov2021}. For instance, the ULX in NGC 5907 exhibited a luminosity about 1000 times the Eddington limit \citep{Israel2017Sci}. Other ULXs, such as M82 ULX-2, M51 ULX-8, and NGC 1313 X-2, have been observed with luminosities approaching $\sim 100$ times the Eddington limit \citep{Grise2008,Bachetti2014,Sathyaprakash2019,Middleton2019,Brightman2020}. Super-Eddington accretion in NS ULXs is generally attributed to strong NS magnetic fields \citep[e.g.,][]{Mushtukov2015,DallOsso2015} and the accretion disk structure (see below). This implies that, despite the relatively short duration of the RLO phase in NS intermediate-mass or high-mass X-ray binaries, it is still feasible for the NS to accrete a sufficient amount of matter and angular momentum, potentially contributing to its recycling.

In a NS + He star binary system,  if the He star is sufficiently massive, it may eventually undergo a supernova  (SN) explosion and form a NS. If this process does not disrupt the binary system, a double NS (DNS) system will form.  In currently observed DNSs, the first-born NSs have spin periods ranging from 17 to 185 ms \citep[See][Table 1]{Deng2024a}, suggesting that they have undergone substantial recycling. Previous studies proposed that recycling of the first-born NSs generally occurs during the Case BB phase in the NS + He star stage \citep{Tauris2013,Tauris2015,Tauris2017,Guo2025}, neglecting accretion before the CE phase. The discovery of PSR J1928+1815 indicates that the latter may also play a role, thus warranting further investigation.

The remainder of the paper is structured as follows: Section 2 introduces the models employed in our study. Section 3 presents the calculated results. Sections 4 and 5 discuss the uncertainties in this work and summarize the results, respectively.

\section{Method and assumptions}
To simulate the binary star evolution, we employed both the  binary population synthesis code \texttt{BSE} originally developed by \citet{Hurley2002} and updated by \citet{Shao2014} and the \texttt{MESA} stellar evolution code \citep[version 11554][]{Paxton2011,Paxton2013,Paxton2015,Paxton2018,Paxton2019}. 
We used the \texttt{BSE} code to model the formation of NS X-ray binaries. We assumed a constant star formation rate of $3\,M_{\odot}$\,yr$^{-1}$ over 12 Gyr. The primordial binaries consisted of a primary star with mass $M_1$ and a secondary star with mass $M_2$ in circular orbits. The primary star's mass distribution followed the initial mass function (IMF) of \citet{Kroupa1993}, spanning $5-40\,M_{\odot}$. We adopted a uniform distribution between 0 and 1 for the mass ratio ($q = M_2/M_1$), as suggested by \citet{Kobulnicky2007}. The initial orbital separation ($a$) was sampled from a logarithmic uniform distribution between 3 and $10^4$ $R_{\odot}$, following \citet{Abt1983}.

Stellar wind mass loss was modeled using the fitting formula of \citet{Nieuwenhuijzen1990}. For OB stars with effective temperatures above 11,000 K and for stripped He stars, we used the simulated relations of \citet{Vink2001} and \citet{Vink2017}, respectively. We assumed that half of the mass transferred from the primary star via RLO was accreted by the companion star and adopted the corresponding critical mass ratios, $q_{\rm cr}$ calculated by \citet[][their Figure 1]{Shao2014} as the criterion for stable MT. In our model, if MT becomes dynamically unstable, a CE phase ensues.

To model CE evolution, we compared the binary's orbital energy to the binding energy of the donor's envelope \citep{Webbink1984}
\begin{equation}
    \alpha_{\rm CE}\left(\frac{GM_{\rm 1c}M_2}{2a_{\rm f}}-\frac{GM_1M_2}{2a_{\rm i}}\right)=\frac{G(M_1-M_{\rm 1c})M_1}{\lambda r_{\rm L}},
\end{equation}
where $\alpha_{\rm CE}$ represents the CE ejection efficiency, $M_{\rm 1c}$  the mass of the primary's core, $a_{\rm i}$ and $a_{\rm f}$ the orbital separations of the binary before and after the CE stage respectively, $r_{\rm L}$ the RL radius at the onset of CE, and $\lambda$ the binding energy parameter of the primary's envelope.  If the orbital energy exceeds the binding energy, the envelope is assumed to be ejected. To assess the potential influence of $\alpha_{\rm CE}$, we considered three values: 0.3, 1, and 3. The binding energy parameter, denoted as $\lambda$, which characterizes the structure of the stellar envelope, is influenced by both the mass of the star and its evolutionary stage \citep{Ivanova2013}. Building on the work of \citet{Xu2010a,Xu2010b}, \citet{Wang2016RAA} used the  \texttt{MESA} code to systematically calculate  $\lambda$ values, adopting a modified definition of the core-envelope boundary. We extended this approach by using \texttt{MESA} to calculate  $\lambda$ values for a denser grid of stellar masses and radii.

For the modeling of NS formation, we incorporated a rapid explosion mechanism to determine remnant masses and natal kicks, following the framework of \citet{Fryer2012}. We used the criterion of \citet{Fryer2012} to distinguish between NSs formed in core-collapse supernovae (CCSN) and electron-capture supernovae (ECSN). NSs formed via CCSN were assigned a kick velocity drawn from a Maxwellian distribution with a standard deviation of $\sigma_{\rm CCSN}=$ 265 km\,s$^{-1}$  \citep{Hobbs2005}; NSs from ECSN received a lower kick velocity with $\sigma_{\rm ECSN}=$ 20 km\,s$^{-1}$ \citep{Podsiadlowski2004,Heuvel2004,Verbunt2017}.

Our calculations indicate that in the resulting NS-main-sequence (MS) binary systems, over 60\% of the NSs originate from ECSN. Following the prescription of \citet{Fryer2012}, the NS mass formed through this channel is uniformly set to 1.38\,$M_\odot$. Accordingly, in our subsequent detailed evolutionary calculations, we adopted an initial NS mass of 1.4\,$M_\odot$ for all systems. However, we note that the NS masses generated by the population synthesis simulations exhibit a distribution ranging from $\sim$1.1 to $\sim$2.0\,$M_\odot$, rather than being entirely concentrated at 1.4\,$M_\odot$.

For the stellar evolution calculations performed with \texttt{MESA}, each binary system was initialized with a NS and a MS companion of Solar metallicity \citep[][Z = 0.0142]{Asplund2009}. We adopted the mixing-length theory of \citet{Bohm1958}, as outlined by \citet{Cox1968}, to model convection. The mixing-length parameter $\alpha_{\rm MLT}$ was set to 2. Convective regions were identified using the Ledoux criterion \citep{Ledoux1947}. We incorporated overshooting from the hydrogen-burning convective core using a step-overshooting scheme, extending the core's size by $\alpha_{\rm ov}$  = 0.335 pressure scale heights in accordance with \citet{Brott2011}. For convective cores after the MS, we included a small degree of exponential overshooting \citep{Herwig2000} with a decay length scale of $f = 0.01$, as overshooting from other convective regions is less well characterized.

We adopted the modified version of the \citet{Kolb1990} method to model MT, incorporating the effects of extended atmospheres as described in \citet{Ritter1988} and considering the possibility of optically thick regions overflowing the donor's RL. The microphysics of accretion imposes several additional constraints on the accretion rate. One such constraint is the Eddington limited accretion. When the accretion luminosity reaches $L_{\rm Edd}=4\pi GM_{\rm NS}c/\kappa$ (where $G$ is the gravitational constant, $M_{\rm NS}$ is NS's mass, $c$ is the speed of light, and $\kappa$ is the opacity assumed
to be due to pure electron scattering), radiation pressure balances gravity, halting steady accretion. This luminosity limit corresponds to an accretion rate limit \citep{Tauris2013}
\begin{equation}
\dot{M}_{\text{Edd}} = 4.6 \times 10^{-8} \, \mathrm{M}_{\odot} \, \mathrm{yr}^{-1} \cdot M_{\text{NS}}^{-1/3} \cdot \frac{1}{1 + X_{\text{H}}},
\end{equation}
where $X_{\rm H}$ is the H mass fraction of the accreted matter. For super-Eddington accretion, \citet{Chashkina2017} extended the standard thin disk model proposed by \citet{Shakura1973} to geometrically thick disk accretion. They argued that when the disk's half-thickness is comparable with the inner radius, the local Eddington limit is reached if the inner radius $R_0$ exceeds the spherization radius $R_{\mathrm{sph}}$. Here $R_0$ is taken to be the NS magnetospheric radius approximated by the Alfv\'en radius
\begin{equation}
    R_0\simeq R_{\rm A}=\left(\frac{\mu^2}{\dot{M}_{\rm in}\sqrt{2GM_{\rm NS}}}\right)^{2/7},
\end{equation}
where $\mu= BR^3_{\rm NS}$ is the magnetic moment of the NS, $B$ is the surface magnetic field, and $\dot{M}_{\rm in}$ is the accretion rate at $R_0$. The spherization radius is
\begin{equation}
    R_{\mathrm{sph}}\approx \frac{3\kappa \dot{M}_{\rm MT}}{8\pi c},
\end{equation}
where $\dot{M}_{\rm MT}$ denotes the mass transfer rate. This yielded the critical accretion rate \citep{Chashkina2017}
\begin{equation}
    \dot{M}_{\rm crit,1}\simeq 35\dot{M}_{\rm Edd}\left(\frac{\mu}{10^{30}\rm \,G\,cm^{3}}\right)^{4/9}.
\end{equation}
At higher accretion rates, advection becomes dominant in the disk \citep{Lipunova1999}, leading to an enhanced critical accretion rate \footnote{The magnitude of $\dot{M}_{\rm crit}$ is contingent upon several additional parameters: the viscosity parameter, the vertical effective polytropic index of the accretion disk, and the energy fraction that powers the disk winds. Here, we adopt the value given by \citet{Gao2022}.} \citep{Chashkina2019}
\begin{equation}
    \dot{M}_{\rm crit,2}\simeq 200\dot{M}_{\rm Edd}\left(\frac{\mu}{10^{30}\rm \,G\,cm^{3}}\right)^{4/9}.
\end{equation}
Accordingly we set the upper limit of $\dot{M}_{\rm in}$ to
\begin{equation}
\dot{M}_{\text {crit }} \simeq \begin{cases}\dot{M}_{\text {crit },1} & \text { if } \dot{M}_{\rm MT}<\dot{M}_{\text {crit}, 2} \\ \dot{M}_{\text {crit },2} & \text { if } \dot{M}_{\rm MT}>\dot{M}_{\text {crit},2}\end{cases}.
\end{equation}
To address the maximum accretion rate ($\dot{M}_{\rm NS,max}$) onto the NS during MT, we followed \citet{Chashkina2017,Chashkina2019} and set $\dot{M}_{\rm NS,max}=\dot{M}_{\rm crit}$.
If the MT rate exceeds the $\dot{M}_{\rm NS,max}$, we postulated that the excess material is expelled from the binary system in the form of isotropic wind from the NS.

To evaluate the impact of the initial magnetic field $B_0$ on the calculated results, we considered three different values: $B_0=10^{11}$, $10^{12}$, and $10^{13}$ G, respectively\footnote{Due to the strong NS magnetic field, the accreted matter is funneled to the relatively small polar areas of the NS. The temperature and pressure are sufficient for stable thermonuclear burning, precluding the appearance of thermonuclear bursts \citep{Bildsten1997}. Thus, we assumed all the accreted matter remains within the NS.}. We adopted an analytical form of magnetic field evolution due to accretion, as given by \citep{Shibazaki1989,Zhang2006}
    \begin{equation}
        B=\frac{B_0}{1+\Delta M/m_{\rm B}},
    \end{equation}
where $\Delta M$ is the accreted mass of NS and $m_{\rm B}\in (10^{-5},10^{-3}) M_{\odot}$ is a model parameter. Here we adopted $m_{\rm B}=10^{-4} M_{\odot}$. 

According to \citet{Tauris2012}, the relation between the final NS spin period and the accreted mass is
\begin{equation}
    P_{\rm s}(\rm{ms})=(\Delta M_{\rm NS}/0.22M_{\odot})^{-3/4}(M/M_{\odot})^{1/4}.
\end{equation}
However, this formula was based on the assumption that the accretion torque is sufficiently strong  to maintain the NS around the equilibrium period. Should the mass accretion timescale be shorter than the spin-up timescale, the NS's spin period will deviate from the instantaneous equilibrium period. Thus, it is necessary to simultaneously follow the evolution of the magnetic field and the spin period.

The evolutionary state of the NS is determined by comparing its light-cylinder radius ($R_{\rm l}=c/\omega$), corotation radius ($R_{\rm co}=[GM/\omega^2]^{1/3}$), and the inner radius of the accretion disk ($R_0$):
\begin{itemize}
    \item the ejector state ($R_{\rm 0}>R_{\rm l}$):  The pulsar wind pressure prevents the companion star's wind from reaching the NS's light cylinder. The NS acts as a radio pulsar;
    \item the propeller state ($R_{\rm co}<R_{\rm 0}<R_{\rm l}$): The infalling matter penetrates into the light cylinder but is halted by the centrifugal force;
    \item the accretor state ($R_{\rm 0}<R_{\rm co}$): The accreting matter is channeled onto the NS magnetic poles following the NS magnetic field lines.
\end{itemize}

The spin evolution of a NS is governed by the angular momentum conservation equation
\begin{equation}
    \frac{dI\omega}{dt}=K_{\rm su}-K_{\rm sd},
\end{equation}
where $I$ is the moment of inertia of the NS, and $K_{\rm su}$ and $K_{\rm sd}$ represent the spin-up and spin-down torques, respectively. We adopted the following prescriptions \citep{Lipunov1992,Beskin1993}:
\begin{itemize}
\item in the ejector state, $K_{\rm su}=0$ and $K_{\rm sd}=2\mu^2/R_{\rm l}^3$;  
\item in the propeller state, $K_{\rm su}=0$, and $K_{\rm sd}=k_{\rm t}\mu^2/R_{\rm A}^3$; 
\item in the accretor state, $K_{\rm su}=\eta\,\dot{M}_{\rm NS}\sqrt{GM_{\rm NS}R_{\rm 0}}$ and $K_{\rm sd}=(1/3)\mu^2/R_{\rm co}^3$. 
\end{itemize}
We set $k_{\rm t}=1$ \citep{Shakura2012} and $\eta=0.8$. In the accretor state, the balance between the spin-up and spin-down torques results in the equilibrium spin period
\begin{equation}
P_{\mathrm{eq}}  \simeq (1.40 \mathrm{~ms})B_8^{6/7}\left(\frac{\dot{M}_{\rm NS}}{0.1 \dot{M}_{\mathrm{Edd}}}\right)^{-3/7}\left(\frac{M_{\rm NS}}{1.4 \mathrm{M}_{\odot}}\right)^{-5/7} R_{13}^{18/7},
\end{equation}
where $B_8$ is $B$ in units of $10^8$ G, and $R_{\rm 13}$ is the NS radius in units of 13 km.

We set the initial NS mass to  $1.4\,M_{\odot}$ and the initial donor mass between 1 and $20\,M_{\odot}$ in steps of 0.5\,$M_{\odot}$. The logarithm of the initial binary orbital period was distributed over the range $-0.5\leq \log P_{\rm orb,i}({\rm day})\leq 4$ in steps of $\Delta\log P_{\rm orb,i}({\rm day}) =0.05$. Following \citet{Shao2021}, we assumed that CE evolution occurs if either of the following conditions is satisfied: (1) $\dot{M}_{\rm MT}>\dot{M}_{\rm trap}=2\dot{M}_{\rm Edd}R_{\rm L,NS}/R_{\rm NS}$, where $R_{\rm L,NS}$ is the RL radius of NS \citep{King1999,Belczynski2008}; and (2) $\dot{M}_{\rm MT}>0.02M_{\rm donor}/P_{\rm orb}$ \citep{Pavlovskii2015}; and $R_{\rm donor}>R_{\rm L_2}$ \citep{Ge2020}. Each binary evolutionary track was terminated if CE evolution occurred or the model numbers exceeded 10,000.

\section{Results}

Figure 1 illustrates an example of pre-CE RLO  MT. The companion star begins to fill its RL at approximately 26.83 Myr. The MT rate progressively increases until 26.91 Myr, when it surpasses $\sim 10^{-2}\,M_{\odot}$\,yr$^{-1}$. This triggers dynamically unstable MT and the subsequent CE evolution (panel a). The pre-CE RLO phase persists for approximately 80, 000 years. A stronger initial NS magnetic field enables a higher accretion rate and greater accreted mass. During the RLO MT, the companion star loses $\sim$ 0.3\,$M_{\odot}$ of its envelope, the orbital period decreases from 891 days to 670 days (panel b), and the He core mass of the companion star increases to 2.5 $M_{\odot}$ (panel c). Due to accretion, the NS magnetic field weakens by one to two orders of magnitude (panel e).

\begin{figure*}
    \centering
    \includegraphics[width=0.9\linewidth]{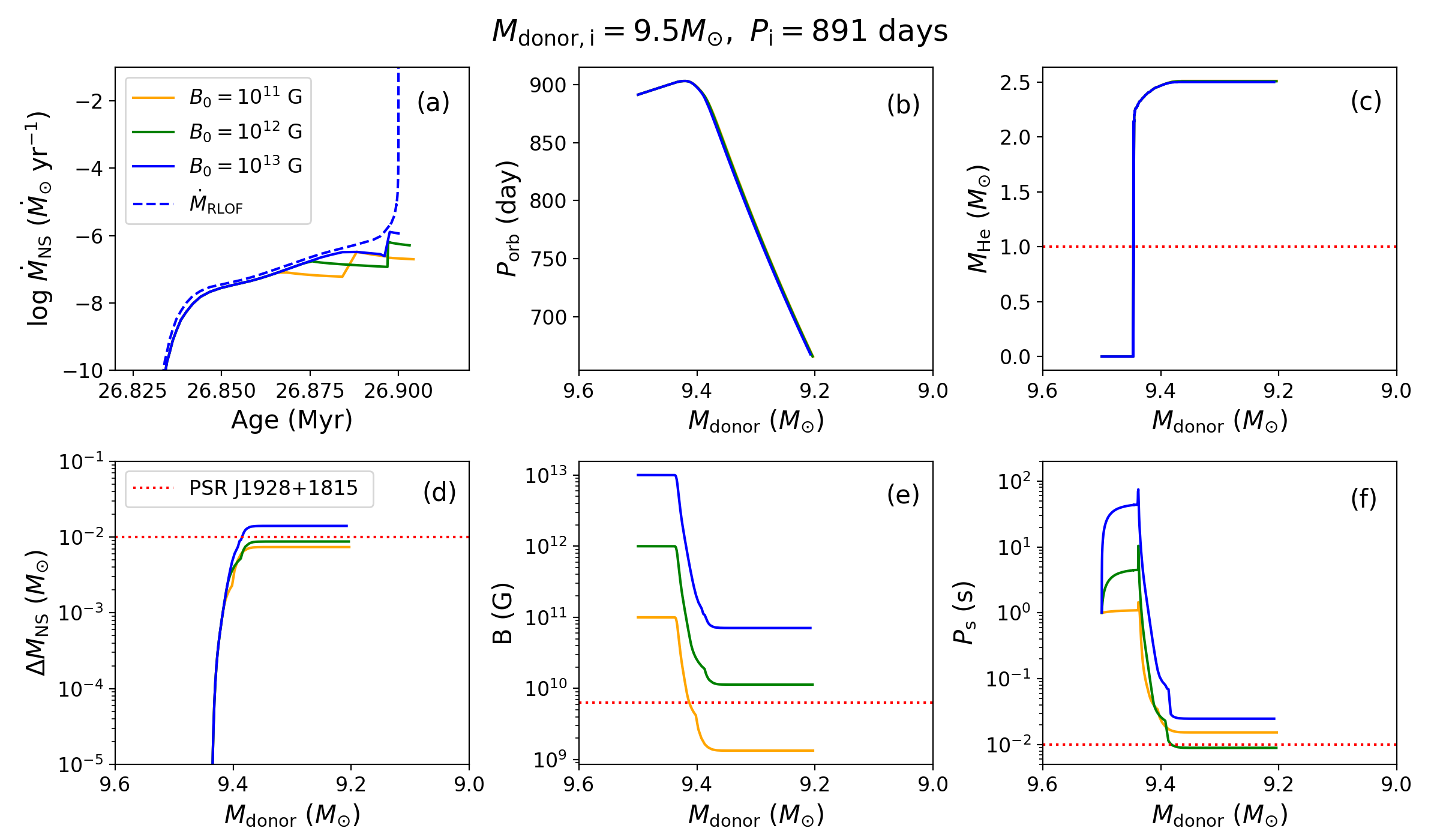}
    \caption{Evolution of a 1.4\,$M_{\odot}$ NS and a 9.5\,$M_{\odot}$ MS donor, commencing with an orbital period $P_{\rm orb,i}$ = 891 days. Panel (a) depicts the evolution of the accretion rate as a function of time. The blue dashed line represents the RLO MT rate for a system containing a NS with an initial magnetic field of $B=10^{13}$ G. Panels (b-f) illustrate the evolution of binary orbital period, donor star's He core mass, NS accreted mass, magnetic field and spin period as a function of the donor mass. The yellow, green, and blue solid lines correspond to initial NS magnetic fields of $10^{11}$, $10^{12}$, and $10^{13}$ G, respectively. The red dashed line denotes the observed values of J1928+1815. Note that the required accreted mass of 0.01 $M_{\odot}$ for accelerating J1928+1815 was derived by applying Eq.~(9). }
    \label{fig:f1}
\end{figure*}

The relationship between the NS's spin evolution and its initial magnetic field (or accreted mass) is non-monotonic (panel f). Prior to the RLO MT, the NS resides in either an ejector or propeller state, leading to spin-down. A stronger initial magnetic field results in a more pronounced spin-down. During the accretion-induced spin-up phase, the NS with an initial magnetic field of $10^{12}$ G achieves the shortest spin period (approximately 8.9 ms), whereas those with initial fields of $10^{11}$ and $10^{13}$ G have spin periods of 15.3 ms and 24.7 ms, respectively. The accreted mass correlates  with the initial magnetic field, attaining values of 0.0074, 0.0087, and 0.0140\,$M_{\odot}$ for the initial magnetic fields of $10^{11}$, $10^{12}$, and $10^{13}$ G, respectively (panel d).

Figure 2 presents the initial companion mass-orbital period distribution for NS X-ray binaries, alongside their corresponding final evolutionary outcomes. Here, we set CE efficiency $\alpha_{\rm CE}=1$, initial NS magnetic field $B_0=10^{12}$ G, and initial spin period $P_{\rm s,0}=1$ s as fiducial parameters. The figure shows that systems with initial orbital periods exceeding $\sim$1000 days typically do not undergo RLO MT (triangles). These wide binaries may evolve into symbiotic binaries when the companion star becomes a red giant star or asymptotic giant star \citep{Lv2012,Yungelson2019,Deng2024c}. Only part of the regions with initial companion masses $\leq 4.5\,M_{\sun}$ undergo stable MT (stars). These systems can typically evolve into partially recycled pulsar binaries \citep{Tauris2000,Shao2012}. Apart from this, all other regions (squares, diamonds, and circles) undergo unstable MT and enter a CE evolution phase. Squares and diamonds represent systems undergoing unstable Case B and Case C mass transfer, respectively, both leading to the formation of NS + He star binaries after successful CE ejection. They can eventually  evolve into NS-WD binaries or DNSs \citep[provided the second supernova does not disrupt the binary;][]{Shao2018a,Vigna2018,Deng2024a}. On the other hand, if the binary systems fail to eject the envelope during the CE phase, the binaries merge (circles), potentially giving rise to Thorne-{\'Z}ytkow objects \citep{Thorne1974,Podsiadlowski1995}.

\begin{figure*}
    \centering
    \includegraphics[width=0.99\linewidth]{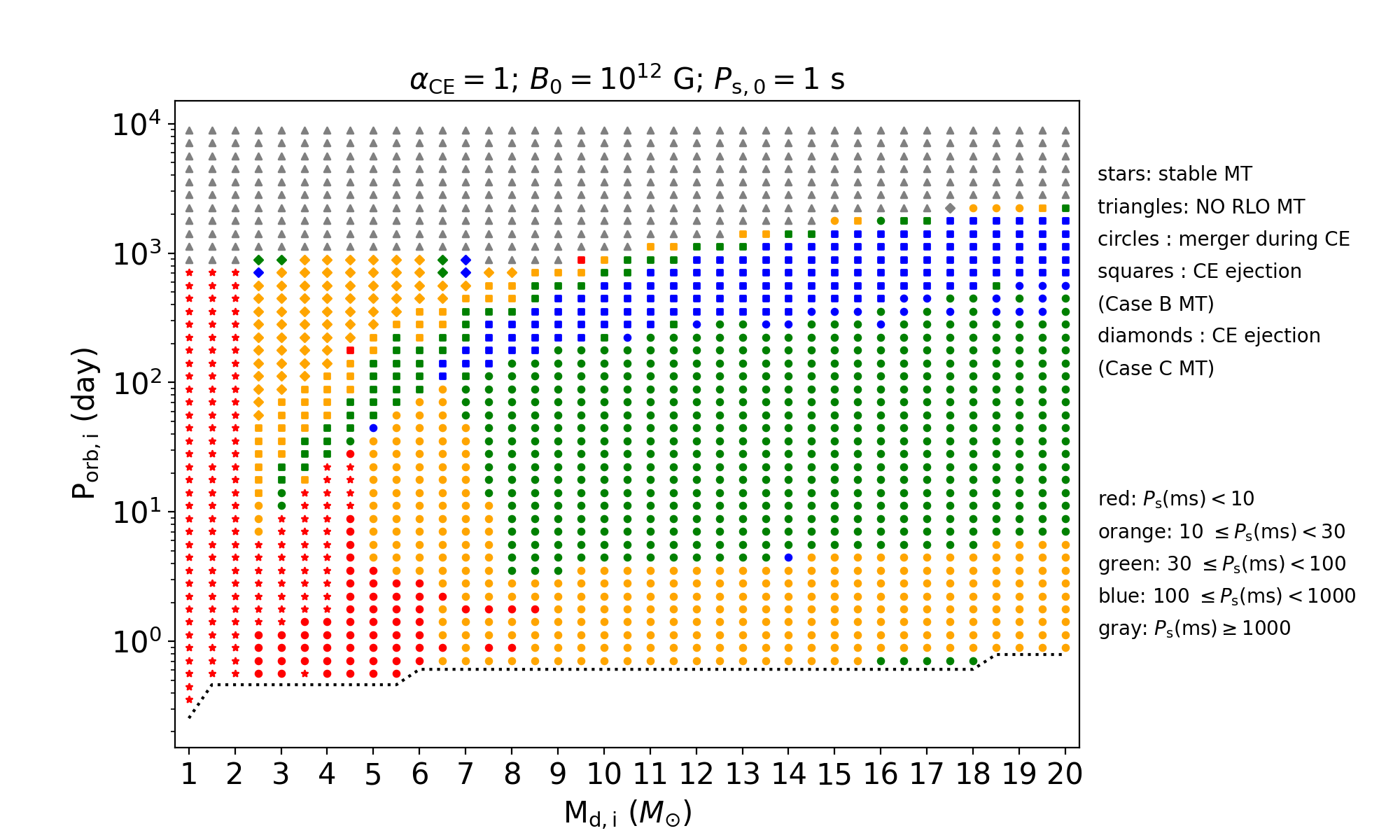}
    \caption{Distinct evolutionary pathways for NS X-ray binaries in the  $M_{\rm d,i}-P_{\rm orb,i}$ plane. Stars denote binary systems capable of stable MT, triangles indicate systems without RLO MT, circles represent systems undergoing unstable MT and entering a CE phase and merging, squares and diamonds signify systems experiencing instable Case B and Case C MT and entering a CE phase but successfully ejecting the envelope, respectively. Red, orange, green, blue, and gray colors correspond to NS spin periods of less than 10 ms, 10-30 ms, 30-100 ms, 100-1000 ms, and greater than 1000 ms, respectively, following the RLO MT phase. The black dotted line marks the orbital period at which the companion star fills its RL at the zero-age main sequence.}
    \label{fig:f2}
\end{figure*}

Focusing on NS + He star progenitor systems, depicted by the squares and diamonds in Figure 2, we find that the parameter space for successful CE ejection  diminishes with increasing initial companion star mass \citep[see also,][]{Gallegos2023,Nie2025}. Only two regions (red squares) yield NS + He star systems with the NS spin periods below 10 ms. The parameter space for the formation of NSs with spin periods between $10-30$ ms, $30-100$ ms, and $100-1000$ ms  progressively expands. This implies that, in case of super-Eddington accretion, pre-CE RLO MT seems able to accelerate the NS spins to the millisecond  regimes \citep[see also][]{Pan2022}.

\begin{figure*}
    \centering
    \includegraphics[width=0.8\textwidth]{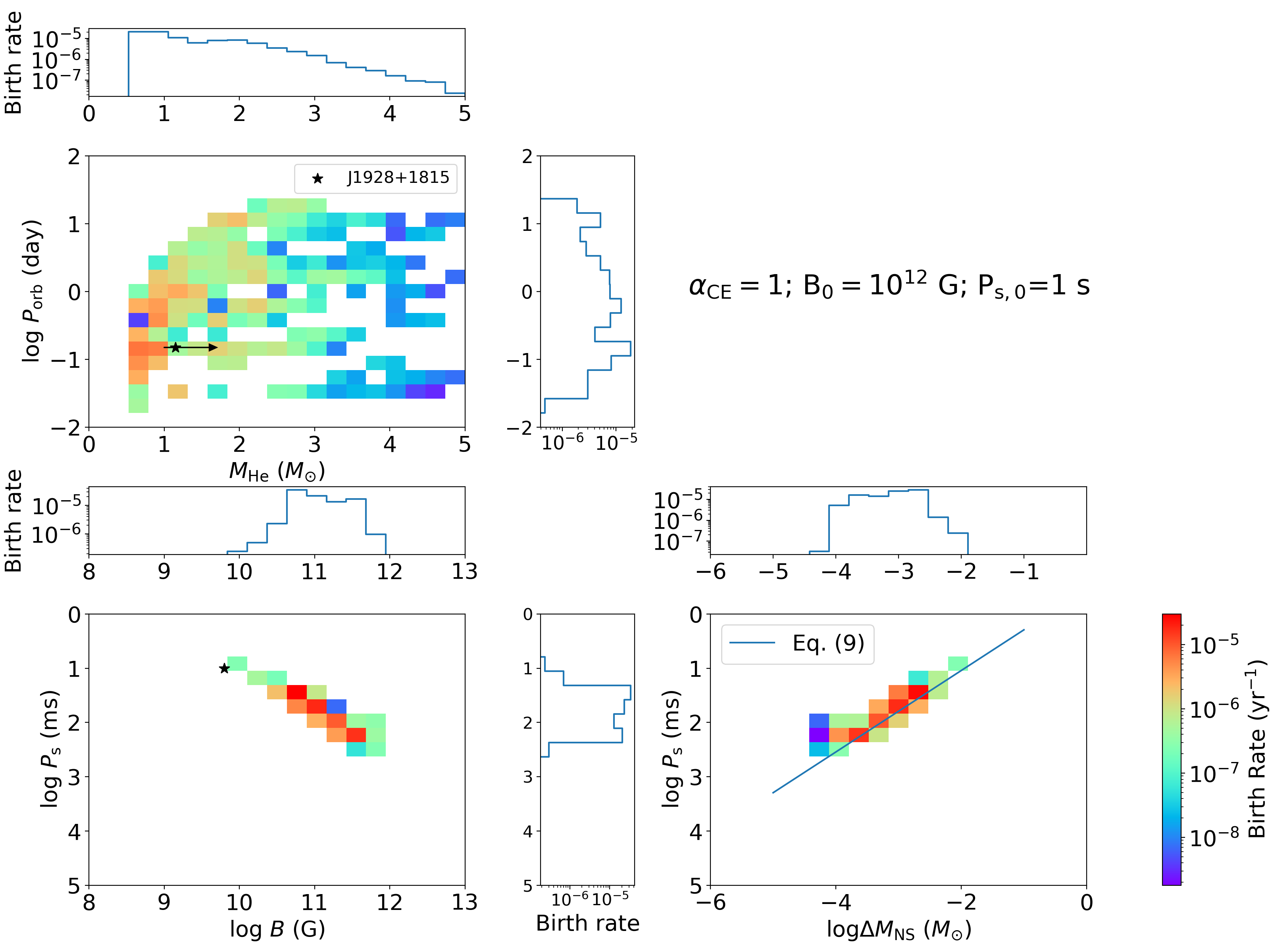}
    \caption{Distributions of NS + He systems  in terms of  companion star mass vs. orbital period, NS magnetic field vs. NS spin period, and accreted mass onto the NS vs. NS spin period (from top left to bottom right). Color intensity represents the birthrate. Here, $\alpha_{\rm CE}$=1, $B_0=10^{12}$ G, and $P_{\rm s,0}=$1 s. In the bottom right plot, the blue line corresponds to Eq. (9). The black star indicates the observational parameters of J1928+1815. }
    \label{fig:f3}
\end{figure*}

Figure 3 shows the distributions of the He star mass, orbital period, NS spin period, magnetic field, and accreted mass for NS + He star systems. We find that the orbital period ranges from less than 1 hour to several tens of days, with most systems concentrated between 0.1 and 1 day; the He star mass spans from 0.6 to 5 $M_{\odot}$, peaking around 1 $M_{\odot}$; the NS magnetic field  varies from $10^{10}$ to $10^{12}$ G, with a peak $\sim 10^{11}$ G; the spin period ranges from $\sim$10 to 300 ms, and the accreted mass spans from $10^{-4}$ to $10^{-2}$ $M_{\odot}$. The spin rate of NSs increases with its accretion of mass (angular momentum), which roughly agrees with Eq.~(9). The simulation results successfully reproduce the magnetic field and spin period of J1928+1815, as well as the He star mass and orbital period of this system.

Figure A.1, analogous to Figure 3, presents results for initial magnetic fields of $10^{11}$ G and $10^{13}$ G. Altering the initial magnetic field has minimal impact on the distributions of He star mass, orbital period, or the overall birthrate of NS + He binaries. However, it significantly affects the final NS magnetic field, spin period, and accreted mass. Reducing the initial magnetic field  to $10^{11}$ G results in weaker final magnetic fields, diminished accreted mass, and broader spin period distributions. Conversely, increasing the initial field to $10^{13}$ G  leads to stronger final magnetic fields, narrower spin period distributions, and a moderate enhancement in the accreted mass.

For a given initial magnetic field, the spin rate of NSs increases with its accretion of mass (angular momentum). Nevertheless,  only an intermediate initial field ($10^{12}$ G) can reproduce the observed properties of systems such as J1928+1815. Higher or lower $B_0$ values fail to simultaneously match both the magnetic field and spin period, suggesting a constrained range for viable formation scenarios.

In comparison, we also consider an initial NS spin period $P_{\rm s,0}=$0.01 s (Figure A.2). The resulting parameter distributions are  virtually identical to those obtained with $P_{\rm s,0} = 1$ s, indicating  that the initial spin period has a negligible effect.

Figure A.3 presents the results for the  CE ejection efficiency $\alpha_{\rm CE}=0.3$  and 3. The CE ejection efficiency predominantly influences the distribution of the orbital periods and the overall birthrate of these systems, with minimal impact on other system parameters. With $\alpha_{\rm CE}=0.3$, the orbital periods cluster around 0.1 day, ranging from approximately half an hour to several days. Increasing $\alpha_{\rm CE}$ to 3 broadens the orbital period distribution, extending from approximately half an hour to a hundred days. The corresponding birthrates of NS + He star systems are $5.2\times 10^{-5}$ yr$^{-1}$, $9\times 10^{-5}$ yr$^{-1}$, and $1.9\times 10^{-4}$ yr$^{-1}$ for $\alpha_{\rm CE}=0.3$, 1, and 3, respectively. Based on the He stars' MS lifetimes and the corresponding birthrate, we estimate the number of NS + He star binaries in the Milky Way to be $\sim 626$, 1383, and 2684, respectively.

\section{Discussion}
The formation of NS + He star binaries encompasses several distinct accretion phases: wind accretion, RLO accretion, accretion during the CE phase, and post-CE He star wind accretion. Owing to the turbulent nature of material flow in wind accretion, the angular momentum of the captured matter may fluctuate between positive and negative values. Over extended timescales, these components largely cancel each other out. Consequently, the net torque exerted on the NS during this phase may be minimal or potentially negligible \citep{Mao2024}. This phenomenon manifests as spin-up and spin-down reversals in wind-fed X-ray pulsars \citep{Bildsten1997, Malacaria2020} and has been corroborated by hydrodynamical simulations \citep{Blondin2012, Xu2019}. Based on these considerations, we disregard the influence of the wind accretion phase in this study.

Similarly, accretion onto compact objects during a CE is commonly assumed to adhere to the spherically symmetric Hoyle-Lyttleton model \citep{MacLeod2015a,MacLeod2017,Hutchinson2024,Rosselli2024}. Under this model, a NS in the CE phase would also fail to efficiently spin up, which explains our omission of spin acceleration effects during this stage. Furthermore, if the NS had undergone a stable accretion-driven spin-up process during the CE phase, its spin axis would align with the post-CE binary orbital axis  \citep{Bhattacharya1991}. However, in the case of PSR J1928+1815, there is a significant misalignment between the spin axis and the orbital axis \citep{Yang2025}. This feature suggests that the accretion process during the CE phase was typically unstable or directionally random \citep{Biryukov2021}. 

Subsequent to the CE phase, the binary will enter a CE decoupling phase (CEDP), during which the He star may retain a residual hydrogen envelope \citep{Nie2025}. MT from this envelope onto the NS could also induce spin-up. Moreover, NS + He star systems can evolve into DNS binaries, potentially involving Case BB mass transfer \citep{Tauris2017,Jiang2021,Deng2024a,Guo2025}. In summary, for NS + He or DNS systems, the initially formed NS will have undergone multiple accretion episodes, rendering the study of its spin evolution particularly intricate and challenging.

Two additional potential formation channels for NS + He star binaries warrant consideration. One channel involves the double-core CE phase \citep{Brown1995,Dewi2006,Andrews2015,Hwang2015}. In this scenario, the initial binary mass ratio is typically near unity, potentially leading to the formation of DNS systems after the second SN, provided the system remains bound. However, the contribution of this channel to DNS formation varies significantly  across different models and assumptions \citep{Vigna2018,Shao2018a,Deng2024a}. A distinctive feature of this channel is that the resulting NS + He star systems have not experienced pre-CE wind or RLO MT, nor hypercritical accretion during the CE phase. Consequently, the binary would host a non-recycled NS.

Another formation channel is accretion-induced collapse (AIC) of a massive WD \citep[see, e.g.,][]{Miyaji1980,Canal1990,Nomoto1991,Freire2014,Wangbo2018}. In this scenario, if the compact object in an IMXB is a massive oxygen-neon-magnesium (ONeMg) WD (close to the Chandrasekhar mass, $M_{\rm Ch}$), accretion onto the WD may supply sufficient material for its mass to exceed $M_{\rm Ch}$. However, rapid rotation might prevent a super-Chandrasekhar WD from collapsing into a NS \citep{Tauris2013aic,Freire2014}. After the subsequent CE phase, the WD may lose angular momentum due to r-mode instability and/or magnetic dipole radiation, potentially leading to collapse into a MSP \citep{Freire2014}. It should be noted  that the formation of NS+He star systems through this channel requires rather stringent conditions (a narrow initial white dwarf mass range, specific accretion rate ranges, etc.), thus also introducing significant uncertainty. Further investigation is needed in the future work.  

The biggest uncertainty in this work pertains to  CE evolution \citep[see][for a review]{Ivanova2013}, which determines the properties of the resulting NS + He binaries. The criteria for CE evolution, which are closely intertwined with MT stability criteria, remain a subject of ongoing debate \citep{Soberman1997,Pavlovskii2015,Ge2020,Shao2021,Deng2024a}. The post-CE evolutionary fate of a binary system depends on the relative magnitudes of the envelope binding energy and the orbital energy \citep{Webbink1984,Nelemans2005,Klencki2021}. However, significant ambiguity surrounds the choice of the core-envelope boundary for calculating the CE binding energy, and different boundary definitions can significantly impact the calculated value \citep{Dewi2000,Dominik2012,Marchant2021,Klencki2021,Gallegos2023}. The CE ejection efficiency, $\alpha_{\rm CE}$, also carries substantial uncertainty. Theoretically, $\alpha_{\rm CE}$ should be less than or equal to unity. Studies of WD binaries suggest the $\alpha_{\rm CE}$ values of $0.2-0.3$ \citep{Zorotovic2010,Davis2012,Nandez2015,Ge2022,Scherbak2022,Zorotovic2022}. However, research on the formation of black hole LMXBs suggests that $\alpha_{\rm CE}$ is expected to be significantly greater than unity for successful envelope ejection and binary survival \footnote{Some studies suggest that if BHs form via failed supernovae \citep{O'Connor2011}, the formation of BH LMXBs can be explained without invoking unrealistically high $\alpha_{\rm CE}$ values \citep{Wang2016MN,Shao2020,Deng2024b}.} \citep{Kalogera1999,Podsiadlowski2003,Kiel2008,Yungelson2008}. Studies of Galactic DNSs also find that $\alpha_{\rm CE}$ needs to exceed unity to match observations \citep{Chu2022,Sgalletta2023}. Some researchers propose that additional, as-yet-unknown energy sources, supplementing the orbital energy, may contribute to envelope removal \citep{Ivanova2002,Podsiadlowski2010}. Currently, fully three-dimensional hydrodynamic simulations of the CE phase in NS HMXBs remain computationally prohibitive \citep[e.g.,][]{Moreno2022,Vetter2024}. 

\section{Summary}
In this work, we investigate NS accretion in intermediate-mass or high-mass X-ray binaries during the RLO phase preceding CE evolution. Motivated by the discovery of nearly ten NS ULXs \citep{Kuranov2021}, we consider super-Eddington accretion-induced field decay and spin-up in the NSs. By combining \texttt{MESA} with the binary population synthesis code \texttt{BSE}, we explore the formation of NS + He star systems and determine the distributions of the orbital period, He star mass, NS spin period, magnetic field strength, and accreted mass.  Our main findings are as follows:

\begin{enumerate}
    \item For fiducial parameters of $P_{\rm s,0}=1$\,s, $B_0=10^{12}$ G, and $\alpha_{\rm CE}=1$, the resulting NS + He star systems exhibit orbital periods ranging from less than 1 hour to tens of days, He star masses between 0.6 and 5 $M_{\odot}$, NS spin periods from 10 to 300 ms, and NS magnetic fields spanning from $10^{10}$ to $10^{12}$ G. These results effectively cover the observed spin period, magnetic field, He star mass, and orbital period of PSR J1928+1815. Given the initial magnetic field of the NS, the spin rate of NSs increases with its accretion of mass (angular momentum), which roughly agrees with Eq. (9). 
    
    \item Variations in the initial NS magnetic field ($B_0$ = $10^{11}$ and $10^{13}$ G) exert negligible  influence on the final distributions of the He star mass and orbital period but significantly affect the distributions of the NS magnetic field, spin period, and accreted mass. Changes in the initial NS spin period have minimal impact on the final distributions of all system parameters.
    
    \item The CE ejection efficiency predominantly affects the orbital period distribution and system birthrate, with little effect on other parameters. Our calculations yield the birthrate and corresponding total number of NS + He star systems ranging from $\sim 5.2\times 10^{-5}$ to $\sim 1.9\times 10^{-4}$ yr$^{-1}$ and from $\sim 626$ to $\sim 2684$, respectively.
\end{enumerate}

\begin{acknowledgements}
We would like to express our gratitude to the editor and the anonymous reviewer for their time, insightful comments, and valuable guidance throughout the review process. This work was supported by the National Key Research and Development Program of China (2021YFA0718500), the Natural Science Foundation of China under grant No. 12041301 and 12121003.
\end{acknowledgements}

\bibliography{a55872-25}{}
\bibliographystyle{aa}

\begin{appendix}
\section{Additional figures}

\begin{figure*}
    \centering
    \includegraphics[width=0.8\textwidth]{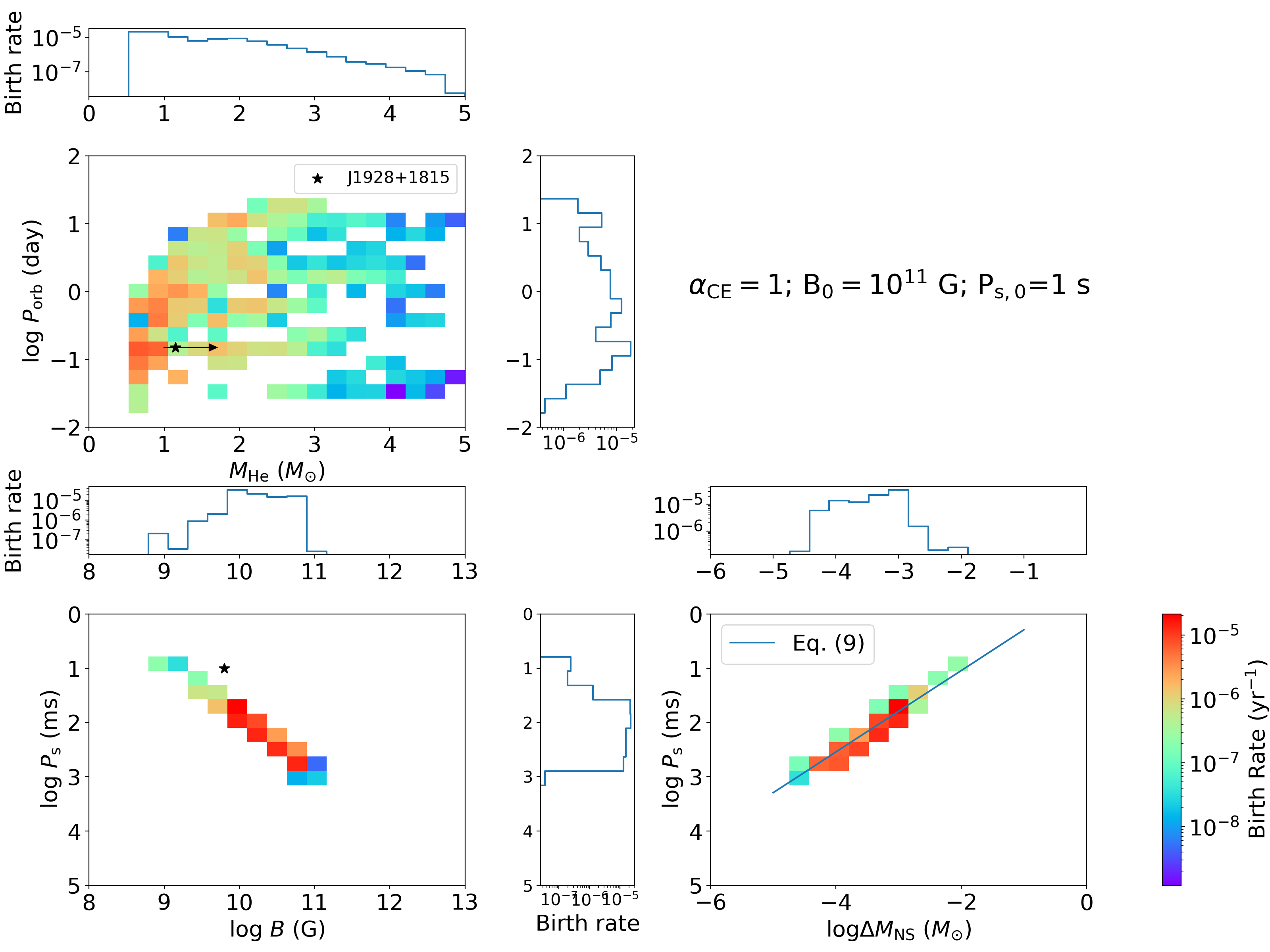}
    \includegraphics[width=0.8\textwidth]{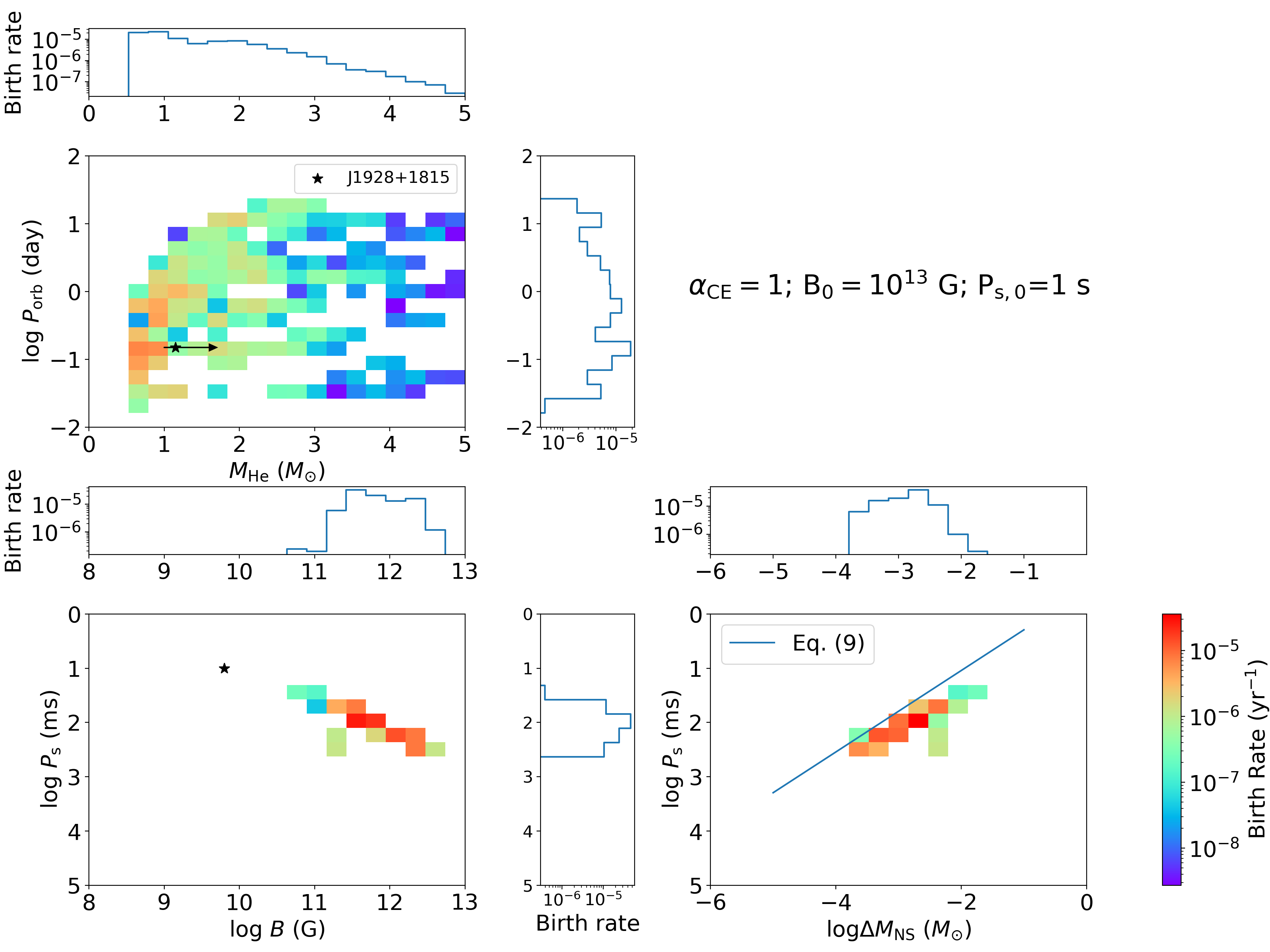}
    \caption{Similar to Figure 3 but initial NS magnetic field $B_0=10^{11}$ G (upper panels) and $B_0=10^{13}$ G (lower panels).}
    \label{fig:fA1}
\end{figure*}

\begin{figure*}
    \centering
    \includegraphics[width=0.8\textwidth]{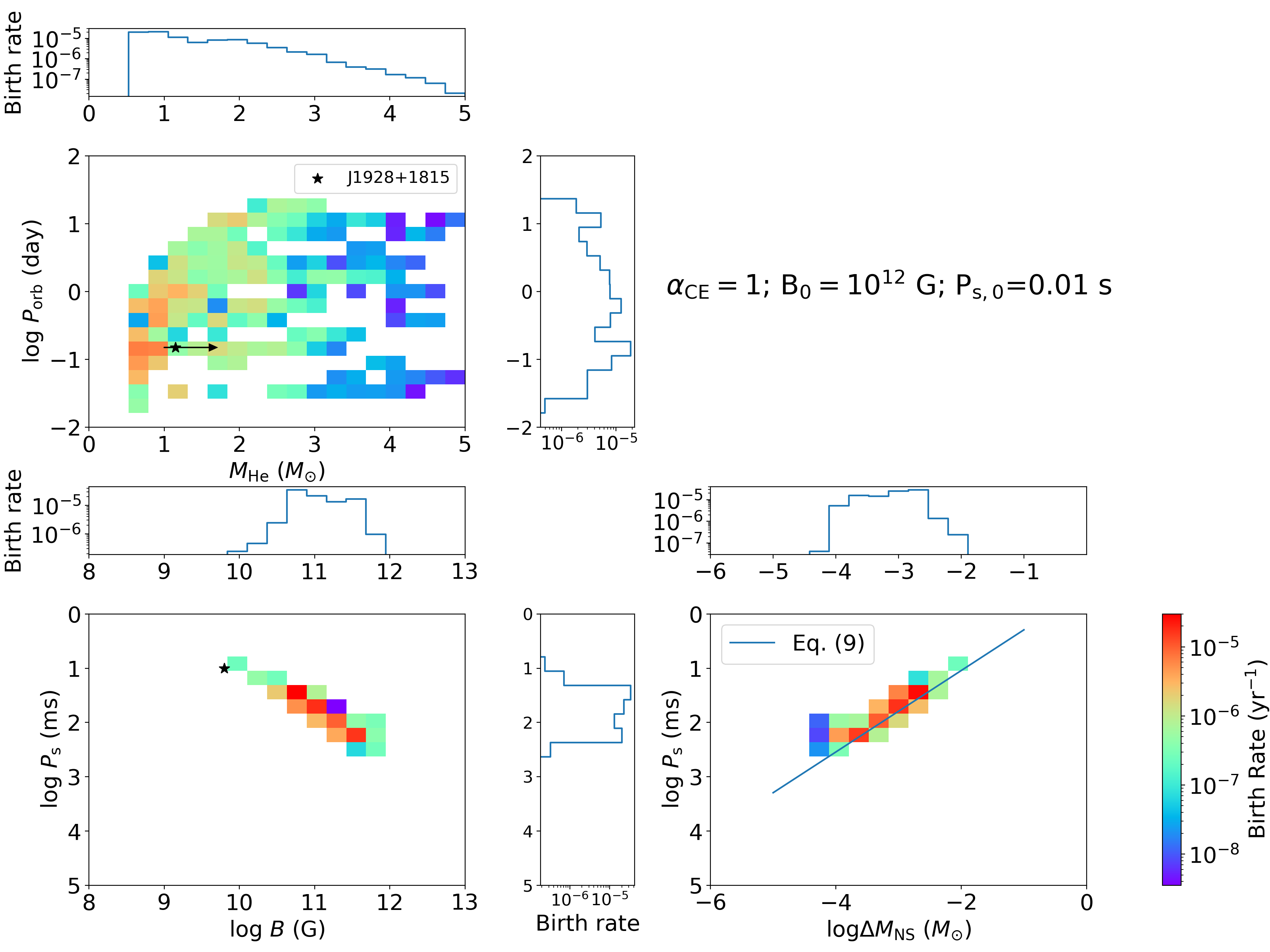}
    \caption{Similar to Figure 3 but for initial NS spin period $P_{s,0}=0.01$ s.}
    \label{fig:fA2}
\end{figure*}

\begin{figure*}
    \centering
    \includegraphics[width=0.8\textwidth]{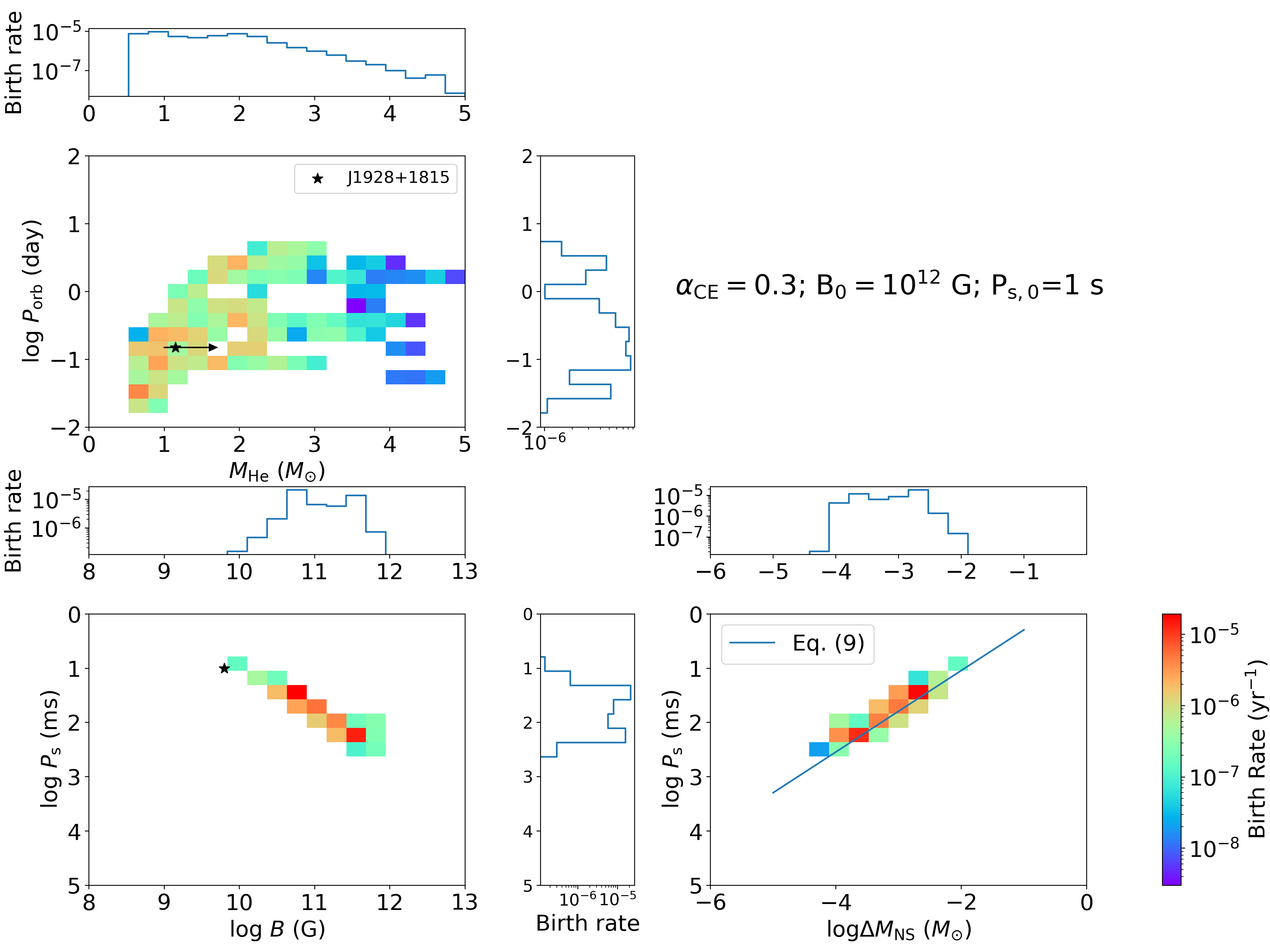}
    \includegraphics[width=0.8\textwidth]{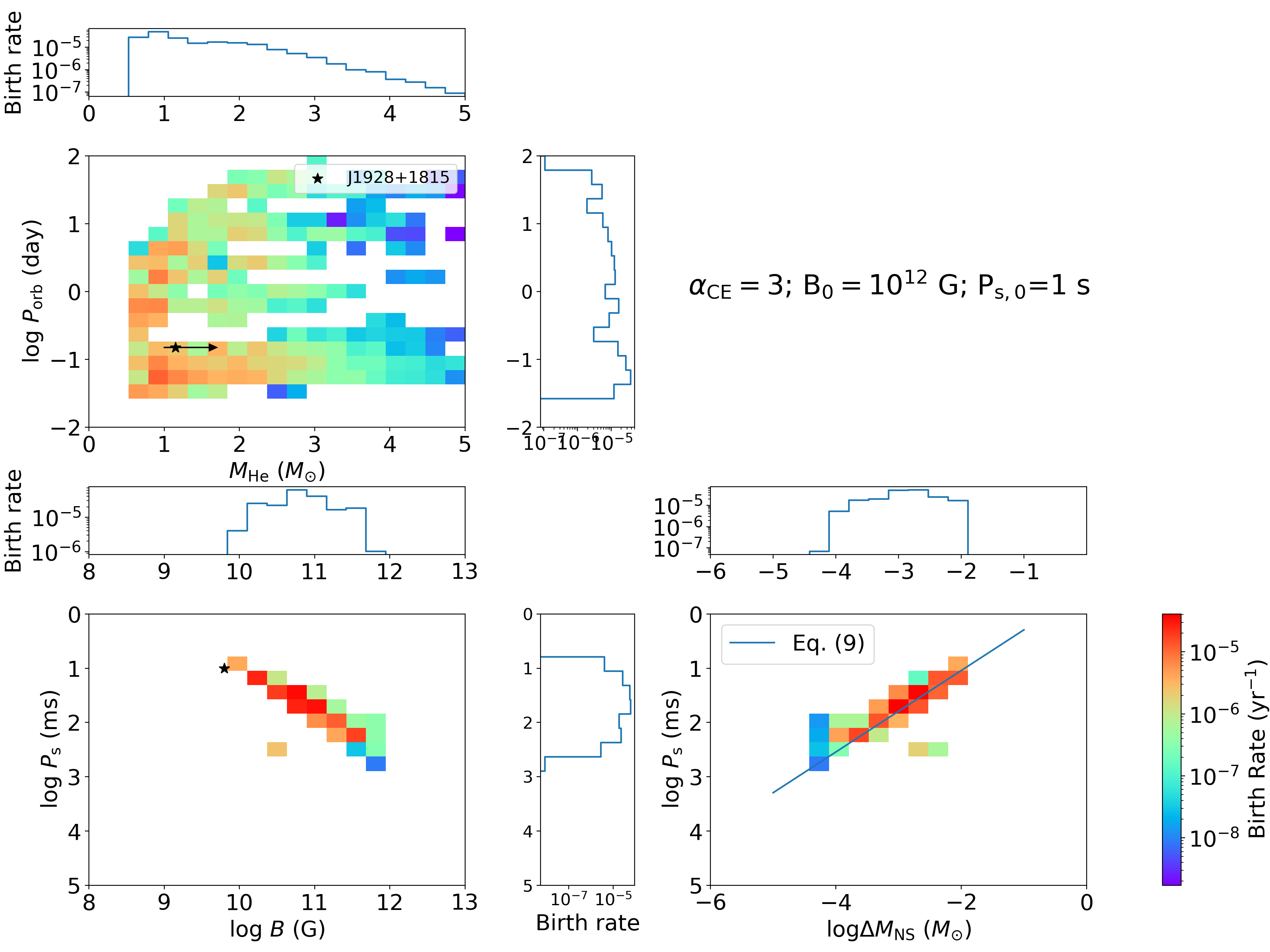}
    \caption{Similar to Figure 3 but for CE ejection efficiency $\alpha_{\rm CE}=0.3$ (upper panels) and $\alpha_{\rm CE}=3$ (upper panels).}
    \label{fig:fA3}
\end{figure*}

\end{appendix}

\end{document}